\documentclass[12pt,draftcls,onecolumn]{IEEEtran}
\IEEEoverridecommandlockouts
\usepackage{cuted}
\usepackage{cite}
\usepackage{bm}
\usepackage{amsmath,amssymb,amsfonts}
\usepackage{algorithm}
\usepackage{algorithmic}
\usepackage{subfig}

\usepackage{graphicx}
\usepackage{makecell}
\usepackage{diagbox}

\usepackage[letterpaper, left=0.625in, right=0.625in, bottom=1.02in, top=0.70in]{geometry}

\usepackage{textcomp}
\usepackage{xcolor}

\newcommand{\be}{\begin{equation}}
\newcommand{\ee}{\end{equation}}

\def\BibTeX{{\rm B\kern-.05em{\sc i\kern-.025em b}\kern-.08em
    T\kern-.1667em\lower.7ex\hbox{E}\kern-.125emX}}

\begin{document}

\title{Precise Coil Alignment for Dynamic Wireless Charging of Electric Vehicles with RFID Sensing }

\author{\IEEEauthorblockN{Haijian Sun\IEEEauthorrefmark{1}, Xiang Ma\IEEEauthorrefmark{2}, Rose Qingyang Hu\IEEEauthorrefmark{2}, Randy Christensen\IEEEauthorrefmark{3}
} \\
\IEEEauthorblockA{
\IEEEauthorrefmark{1}School of Electrical and Computer Engineering, University of Georgia, Athens, GA, USA \\
\IEEEauthorrefmark{2}Department of Electrical and Computer Engineering, Utah State University, Logan, UT, USA \\
\IEEEauthorrefmark{3}Blue Origin LLC, Kent, WA, USA \\
Emails: {hsun@uga.edu, xiang.ma@usu.edu, rose.hu@usu.edu, rchristensen@blueorigin.com}}}

\maketitle

\begin{abstract}
Electric vehicle (EV) has emerged as a transformative force for the sustainable and environmentally friendly future. To alleviate range anxiety caused by battery and charging facility, dynamic wireless power transfer (DWPT) is increasingly recognized as a key enabler for widespread EV adoption, yet it faces significant technical challenges, primarily in precise coil alignment. This article begins by reviewing current alignment methodologies and evaluates their advantages and limitations. We observe that achieving the necessary alignment precision is challenging with these existing methods. To address this, we present an innovative RFID-based DWPT coil alignment system, utilizing coherent phase detection and a maximum likelihood estimation  algorithm, capable of achieving sub-10 cm accuracy. This system's efficacy in providing both lateral and vertical misalignment estimates has been verified through laboratory and experimental tests. We also discuss potential challenges in broader system implementation and propose corresponding solutions. This research offers a viable and promising solution for enhancing DWPT efficiency.
\end{abstract}

\newpage

\section{Introduction}

The past few decades have witnessed tremendous mileage growth in daily commute and freight transportation, fueled by factors such as urbanization, population growth, and economic development. For example, in the United States alone, more than 11 billion tons of freight are transported in over 3 billion miles annually \cite{tamor2022electrification}.  Modern transportation is still mainly powered by gasoline-based internal combustion engines, creating significant impacts across various aspects of socioeconomic and environment. By 2030, the United States aims to achieve a minimum of 50\% electric vehicle (EV)  among all vehicle sales and reach net-zero greenhouse emissions (GHG) by 2050. Consequently, EV has emerged as a transformative force in the automotive industry and broader transportation sector. EVs are crucial for reducing GHG, mitigating climate change, and transitioning towards a sustainable and environmentally friendly transportation future. Besides, EVs can use alternative energy from renewable sources like solar and wind, which can further reduce the dependence on fossil fuels. 
However, there exist significant challenges to the widespread adoption of EVs, primarily centered around batteries (limited energy density, manufacturing, and disposal), range anxiety, and supportive charging infrastructure. 

As a result, today's EVs have to carry heavy battery packs and charge frequently. While there are innovative approaches to alleviate the above challenges, such as battery swapping and new material developed for extremely fast charging and battery, to date, the primary methods to charge EVs fall into either wired or wireless ones. 

\subsection{Wired charging for EVs}
Currently, EV charging is predominated by plug-in chargers located at home or public charging stations. Due to its convenience and lower energy cost, home charging has become a very popular option for EV owners. It is worth noting that the wired EV charging standard varies depending on power type (AC or DC), connector, grid integration, and safety requirements. A non-exhaustive list of standards includes the Society of Automotive Engineers (SAE) J1772 in the United States, Japan, and Australia, International Electrotechnical Commission (IEC) 62196, GuoBiao GB/T 20234 in China, etc. 

Wired or conductive charging has four levels \cite{10102467}. Specifically, Level 1 uses a single-phase 120V AC power that delivers the slowest charging speed (up to 1.92 KW) but usually does not require additional charging infrastructure.  Level 2 can deliver up to around 20 KW thanks to higher voltage and current input, up to 240 V and 80 A, respectively. Although requiring additional fast chargers and professional installation, Level 2 is now commonly available at residential and commercial stations. On top of this, Level 3 defines even higher three-phase voltage and current for achieving 350 KW charging capability. Lastly, Level 4 is a promising technique for substantially decreasing the time for a full charge at the cost of new converters (AC-DC, DC-DC) and highly capable onboard components. 

\subsection{Wireless charging for EVs}
Wireless charging technology, often referred as wireless power transfer (WPT), eliminates the need for physical cables, thus offering a more convenient and flexible approach to recharging EVs. Existing WPT utilizes the coupling effect of electromagnetic fields, including different forms like inductive, resonant, and capacitive coupling for transferring energy over the air. Typically, the WPT system has the transmitter (primary) coil located at the road surface or integrated within the pavement and the receiver (secondary) coil mounted at the bottom of EVs. After the DC-DC converter, energy from the EV-side coil will be fed into the battery. In general, WPT can be further classified into three categories. 
\subsubsection{Stationary WPT (SWPT)} 
SWPT is well-suited for applications in parking lots, fleet depots, and other areas where EVs remain parked during the charging process. SWPT can achieve over 95\% overall efficiency in a 270 KW prototype developed by the Oak Ridge National Laboratory.  As a major first step beyond wired charging, SWPT has been deployed in selected sites backed by car manufacturers, governments, and third-party companies. One example is the company WiTricity, which helped the transition to EV wireless charging in various scenarios.   

\subsubsection{Dynamic WPT (DWPT)} 
\begin{figure}
    \centering
    \includegraphics[width=0.9\linewidth]{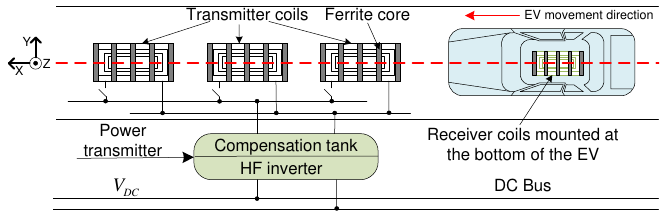}
    \caption{Illustration of a dynamic wireless charging system \cite{10261276}}
    \label{fig:dynamic}
\end{figure}

As the name implies, DWPT involves charging EVs while in motion. Since its inception, DWPT has attracted great interest and shown its potential for 1) extending EV range while reducing battery size and weight, 2) reducing infrastructure complexity and deployment cost, 3) enhancing energy efficiency, and 4) minimizing downtime for public and freight transportation. 
Expanding from SWPT, the implementation of DWPT involves hundreds of primary charging coils connected and spanning miles of range on the road, becoming part of the major infrastructures in future highway systems. 
As shown in Fig. \ref{fig:dynamic}, primary coils, or pads, can be configured separately controlled or formed in groups by roadside unit (RSU). The power transmitter feeds energy into a high-frequency (HF) inverter and compensation tank, then goes to the Ferrite core.  Active communication between RSU and EV is required for authentication, billing, and dynamic charging control \cite{10261276}.  

\subsubsection{Quasi-dynamic WPT (QDWPT)} This concept sits between SWPT and DWPT that provides intermittent wireless charging to EVs during stops or pauses in their operation, such as during brief parking periods or at traffic intersections. When infrastructure is ready, major intersections are expected to deploy the QDWPT system for convenient recharging.  With high output power, even sporadic QDWPT can help alleviate range anxiety.

To summarize, electromagnetic induction is the core of SWPT, DWPT, and QDWPT. Their difference lies mainly in the state of the EV during the charging process. From the implementation perspective, DWPT is more challenging and costly but promising. 
A few large-scale DWPT centers have been established for prototype development and field verification. For example, the ASPIRE center at Utah State University, the FABRIC site at EU, the dynamic EV bus in Korea, and the CIRCE project in Spain.

\subsection{Coil alignment and  technical challenges}
Since the amount of transferred power is determined by the mutual inductance between primary and secondary coils, the energy transfer efficiency of WPT relies on precise alignment between these coils. As illustrated in Fig. 1, misalignment can happen in three directions, namely the longitudinal ($x$), lateral ($y$), and vertical ($z$). Under typical circumstances, lateral misalignment is dominant and cannot be easily adjusted from the coil design perspective. Therefore, most works focus on lateral mitigation. For SWPT and QDWPT, drivers can gradually adjust EV positions following visual or in situ detector assistants to maintain sufficient WPT efficiency. However, with DWPT, the high mobility of EVs makes position alignment very challenging, especially targeting real-time estimation and control. For example, consider a typical WPT charging pad of 1.5 meters length, with the EV traveling at 72 km/h. The total time for contact with the coil is 75 ms. Many technical issues have to be resolved in various aspects, such as 1) real-time positioning and misalignment estimation, 2) real-time adaptive feedback and control, 3) scalability and cost-effective deployment, and 4) resilience and safety considerations. 

\section{Related Works}

As a well-known issue in WPT, coil misalignment has received extensive research attention. Existing works have investigated not only coil misalignment detection and estimation but also mitigation for such issues by designing innovative misalignment-resilient coils.  While the latter is not the focus of this paper, it is worth noting that a rich research community is on it. Their recent works have been summarized in \cite{vu2022operation}. Generally speaking, only a misalignment of no more than roughly 5\% of car width ($\sim$10 cm) can reach 90\% coupling efficiency. The efficiency will quickly diminish when passing that threshold, as specified by SAE J2954. Such accuracy cannot be achieved with GPS systems. In what follows, we briefly introduce related works on existing WPT coil alignment techniques. 

\subsection{Magnetic field-based approach}
Extended from WPT, it is possible to measure the magnetic field values when the EV passes through the transmitter coil to obtain precisely how deviated it is in terms of the relative lateral position. Specifically, one can deploy two additional sensing coils at the left and right sides of the charging pad. Magnetic measurements from these two sensing coils should be equal if well-aligned, i.e., EV passes the charging pad right through the center. Otherwise, one will be larger than another one, a direct signal for lateral misalignment. In \cite{tavakoli2021ev}, they propose to install four additional sensing coils on EVs, and by calculating the voltage values induced in sensing coils, misalignment in both lateral and vertical directions can be derived. To accelerate the estimation process, they have further utilized artificial neural networks (ANN) for post-signal processing. Experimental results on the 500 W DWPT platform have shown $\pm 15 $ cm and $[16.5, 22.5]$ in lateral and vertical estimation accuracy, respectively.

\subsection{Vision-based approach}
Another idea is to use a vision-based method. EVs manufactured in recent years are usually equipped with camera arrays for enhanced driving experience, such as lane keep assist and adaptive cruise control. A natural combination is to facilitate coil alignment with these cameras. This will require coordinated efforts from the very beginning. For example, at the construction stage, use a special paint that marks the position underneath the transmitter coils. In the FABRIC test site,  a grid alignment assistant system (GAAS) was developed to calculate the misalignment between transmitter and receiver coils in both lateral and vertical directions, almost in real-time \cite{8966379}. GAAS targets keeping the EV as close to the center of the coil as possible and achieves a localization accuracy of 20 cm with cheap sensors. Besides, a human-machine interaction display was used to guide the driver with the correct alignment information. Similarly, in \cite{tian2020vision}, \emph{K}-means clustering algorithm is used for processing camera input, which helps identify left and right lane boundaries to determine the center of the coil.

\subsection{Wireless sensing-based approach}
Wireless signals have been widely used for localization. Due to signal randomness caused by multipath and interference, this method usually cannot achieve fine-grained accuracy.  For example, the Wi-Fi-based approach achieves localization accuracy in the sub-meter range. On the other hand, Bluetooth localization can be facilitated by many anchor points but can only reach around 50 cm accuracy \cite{9081910}.  Dedicated wireless devices, such as the ultra-wideband (UWB), use precise time of arrival to obtain sub-10 cm localization accuracy,  a desirable feature for DWPT. In \cite{lee2021ultrawideband}, EV location is calculated concerning two UWB anchor points placed in fixed positions. On EV, two UWB tags are used for the vehicle pose estimation. Experimental results show this method can reliably obtain sub-10 cm localization accuracy.

\subsection{Machine learning-based approach}
Machine learning essentially learns the relationship between input data and output decisions, and in most cases, such a relationship is not explicit. Machine learning has found wide application in various sectors, particularly in computer vision and large language models. Recently, some works have applied machine learning in EV coil alignment. One example is  \cite{chen2023recognizing}. The authors proposed a data-driven strategy to recognize lateral, longitudinal, and yaw angle misalignment. Combined with an orientation-sensitive detection coil, the system can get sufficient measurements via coupling effects and feed into the improved residual network (ResNet) for post-processing. In a 6.6 KW lab prototype, this learning method can achieve less than 1.7 cm localization in 95.2\% cases, showing a high potential. 

\subsection{Voltage detection-based approach}
Misalignment can also be detected by measuring the voltage in the EV's load coil, similar to the magnetic field method. The first step is to analyze the power loss due to misalignment, where the induced voltage can then be derived.  Unlike magnetic mutual inductance, the main challenge of using voltage is to determine whether the load coil is misaligned to the left or the right side. In \cite{hwang2017autonomous}, one sensor coil is placed within the load coil to distinguish the left and right sides via voltage phase difference.  With a voltage comparator, the lateral position estimator can obtain an exact misaligned position, which is then sent to the autonomous coil alignment system (ACAS) for the automatic steering control. Simulation results show an overall accuracy of around 50 cm. 

\subsection{Pros and cons}
The magnetic field and voltage-based approach have proved to be reliable for both accuracy and latency. However, it needs extra sensing components and a modified transmitter coil. Hardware cost and installation may prohibit their wide adoption. 
The vision-based approach is intuitive, simple, and low-cost. It has the potential to be integrated into future EV-driving assistant systems. However, this method needs markers on the road, requiring extra regulations from infrastructure providers.  The primary challenge arises in various adverse conditions, including heavy rain, dense fog, road reflections, snow or ice coverage, or indistinct lane markings. In these scenarios, the effectiveness of the misalignment estimation system can be substantially diminished. Even under ideal conditions, this method can hardly achieve alignment accuracy of sub-10 cm. The wireless signal-based approach has the advantage of being ubiquitous but is susceptible to interference hence not robust in non-line-of-sight (NLOS) scenarios. Besides, deploying a large amount of wireless sensors is not only costly but also needs frequent maintenance. Lastly, the machine learning technique will be more popular as a general framework. Uncertainty to new deployment environments (generalizability) and high computing power requirement are their drawbacks. Table \ref{tab:my_label} summarizes related works. 

\begin{table}
    \centering
    \begin{tabular}{| p{20mm} | p{24mm} | p{32mm} | p{27mm} | p{45mm} |}
    \hline
       \textbf{Related Works} & \textbf{Approach} & \textbf{Alignment Precision} & \textbf{Advantages} & \textbf{Drawbacks} \\
        \hline
        \cite{tavakoli2021ev} & Magnetic field-based &   [-15, 15] cm laterally,  [16.5, 22.5] cm vertically   & reliable, sensitive to misalignment & need dedicated sensing coil; bulky size \\ 
        \hline
        \cite{8966379} & Vision-based & 10 - 20 cm laterally & low-cost, near real-time  &  susceptible to weather and light condition, need marker on the road \\
        \hline
       \cite{lee2021ultrawideband}  & Wireless signal-based & sub-10 cm with UWB, sub-meter with Wi-Fi & low-cost, ubiquitous  & propagation randomness,  sensitive to dynamic environments\\
         \hline
         \cite{chen2023recognizing} & Machine learning and sensing coil & 1.7 cm in simple prototype & near real-time, accurate & lack field experiments and  generalizability \\
        \hline
         \cite{hwang2017autonomous} & Voltage-based & 50 cm in simulations & reliable, sensitive to misalignment & cannot distinguish left or right side misalignment, need infrastructure\\
         \hline
    \end{tabular}
    \caption{Summary of related works}
    \label{tab:my_label}
\end{table}

\section{Enabling Wireless Technology for Precise Coil Alignment}
In this section, we present the proposed RFID-based coil alignment system, which falls into the category of wireless sensing. Radio-frequency identification (RFID) has found a wide range of applications in supply chain, access control and security,  and retail management. An RFID system comprises two main components: the reader and tag. The reader sends RF signals to query nearby RFID tags and energize them. The tag responds with its unique ID via RF backscatter signals. 
Usually the tag keeps a very simple design containing an antenna, microchip, and memory. The passive RFID tag does not have an embedded power source, all energy comes from reader's RF signals. 

\subsection{Advantages of using RFID in DWPT coil alignment}
A single travel time from the reader to the tag and back to the reader can infer their distance information. In fact, utilizing RFID in WPT coil alignment has been found in \cite{shuwei2014research}. It is assumed that the tags are placed on top of the primary coil, and the EV carries an RFID reader to determine its relative distance with tags (coils) to perform coil alignment estimation. We list some advantages for using RFID.

\begin{itemize}
    \item Easy maintenance. RFID tags can be designed to withstand harsh environmental conditions, making them durable for use in various industries. Besides, passive RFID does not need an internal power source, reducing the effort of replacing the batteries. 
    
    \item Low cost. Passive RFID tags usually cost a few cents each, making them a cost-effective solution for large-scale road deployment. 

    \item High performance. RFID systems can provide real-time tracking and localizing of tags. One example is the widely used Impinj reader that can interrogate 1,100 times per second, reducing the  reading latency to less than 1 ms. 

    \item High versatility. The RFID reader can interact with multiple tags simultaneously, thus obtaining a more reliable localization result due to diversity. Besides, the communication range of RFID is usually limited to several meters, which is actually an advantage due to limited interference. 
\end{itemize}

\subsection{Principle of RFID localization} 
Wireless signal-based localization relies on the principle of measuring the characteristics of wireless signals to determine the location of a device or object in space. When the distance between the transmitter and receiver changes, RF parameters such as received signal strength (RSS), phase, and delay will also vary. 
By incorporating the RF physics propagation profile, their distance can be calculated. As mentioned before, the grand challenge is to tackle signal randomness and dynamic environment, particularly the RSS value. In general, phase measurement-based localization proves to be more reliable and accurate.

\begin{figure}[ht]
  \centering
  \subfloat[Principle of RFID localization \cite{yang2014tagoram}]{\includegraphics[width=0.48\textwidth]{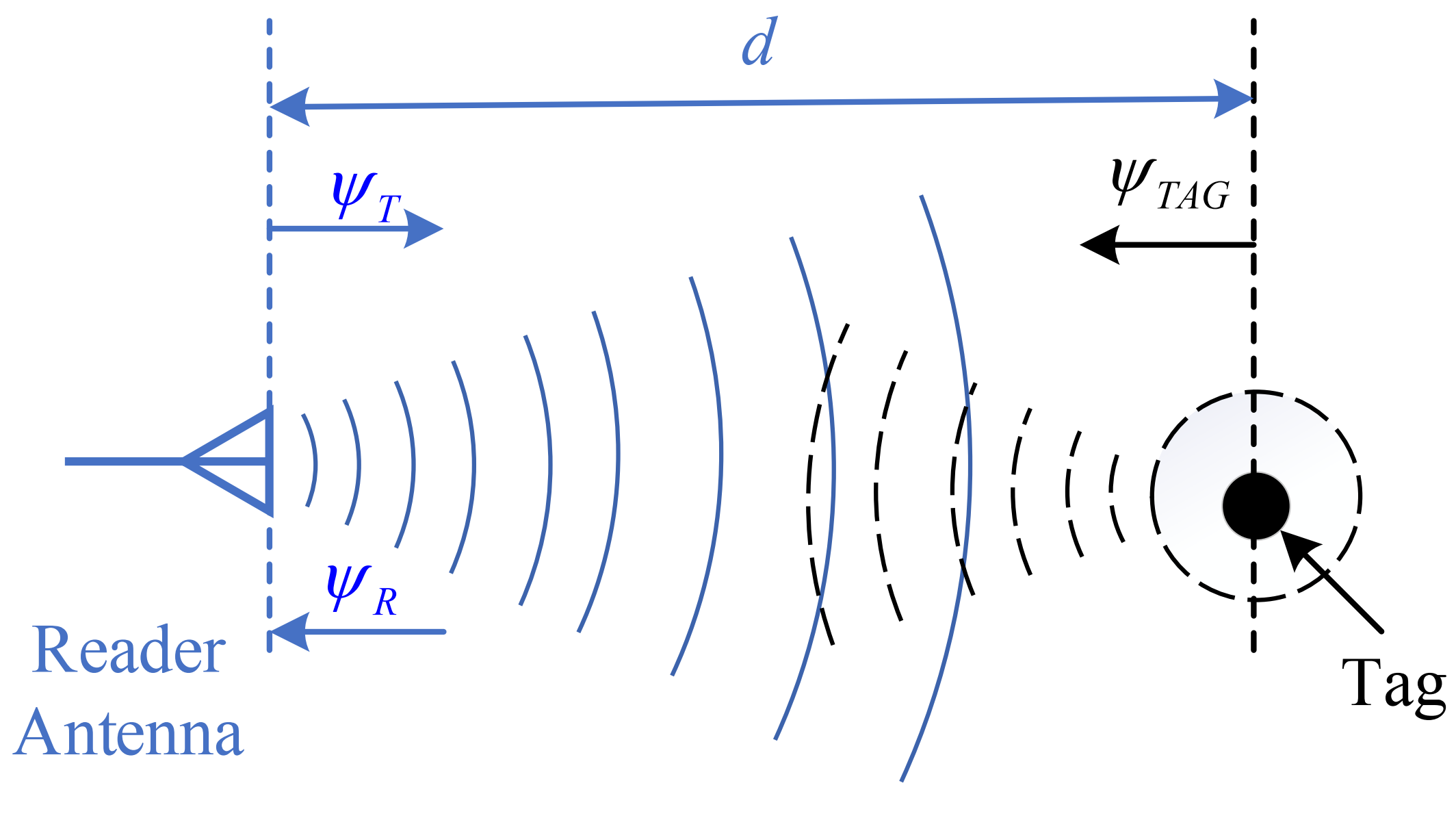}\label{fig:tag}}
  \hfill
  \subfloat[Initial position estimation based on MLE]{\includegraphics[width=0.39\textwidth]{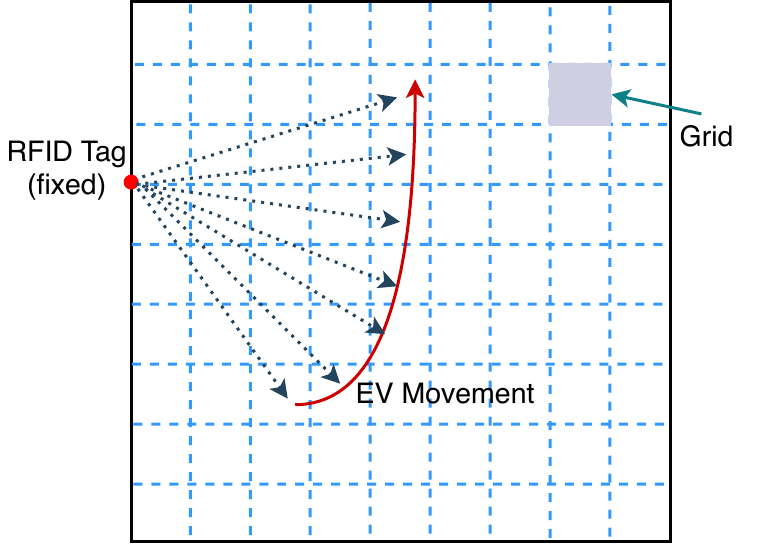}\label{fig:mle}}
  \caption{Brief illustration of RFID localization principle and the MLE algorithm used in this paper.}
  \label{fig:principle}
\end{figure}

As shown in Fig. \ref{fig:tag}, the reader sends a signal to the tag and immediately being reflected back. Ideally, when RF signal travels distance $2d$,  a phase change of $\frac{2d \cdot 2 \pi }{\lambda}$ happens, $\lambda$ is the signal wavelength. Therefore, by measuring the phase difference between transmitting and receiving RF signals, $d$ can be determined. In practice, the reader side observes a received phase $\Psi_r = (\Psi_T + \frac{4 \pi d}{\lambda} + \Psi_R + \Psi_{TAG}) \ mod \ 2 \pi$, where $\Psi_T$, $\Psi_R$, $\Psi_{TAG}$ are phase variances caused by transmit antenna, tag reflection characteristics, and receiver antenna, respectively. The reader can detect $\Psi_r$ and $\Psi_R$. But the other two phase values appear to be random variables and are tag-dependent. Additionally, modulo operation leads to phase ambiguity. These two factors make the exact calculation of $d$ very challenging \cite{yang2014tagoram}. 

\subsection{Maximum likelihood estimation algorithm and control}
Here, we introduce a promising technique for simultaneously resolving random phase values and ambiguity. The key idea is to obtain more coherent $\Psi_r$ measurements such that randomness impact becomes minimum and the phase change should be mainly caused by the distance $d$ \cite{yang2014tagoram}. When the random variables follow Gaussian distribution, maximum likelihood estimation (MLE) is one of the optimum estimators. 


Briefly, as shown in Fig. \ref{fig:mle}, the MLE procedure can be described as follows. 1) The observed area is partitioned into grids, with grid size as a  parameter to be designed. Note that a smaller grid size will result in better resolution but a higher computation cost. Here, grid size is set at \emph{mm}-level for accurate localization.  2) With tag location known in one grid, the theoretical received phase can be calculated at all other grids. 3) Assume the EV moves at a known trajectory. The MLE algorithm calculates the likelihood of the starting position at each grid, essentially creating a moving synthetic aperture radar (SAR) in a coherent way. This joint estimation will effectively compress the phase randomness. 


\section{Trail Experiments and Field Test}
In this section, we first show the system architect and then present our experiment setup in both indoor track and outdoor field experiments, followed by their respective results. 

\subsection{System architect and design} 
\begin{figure}[ht]
  \centering
  \subfloat[Lab experiments]{\includegraphics[width=0.487\textwidth]{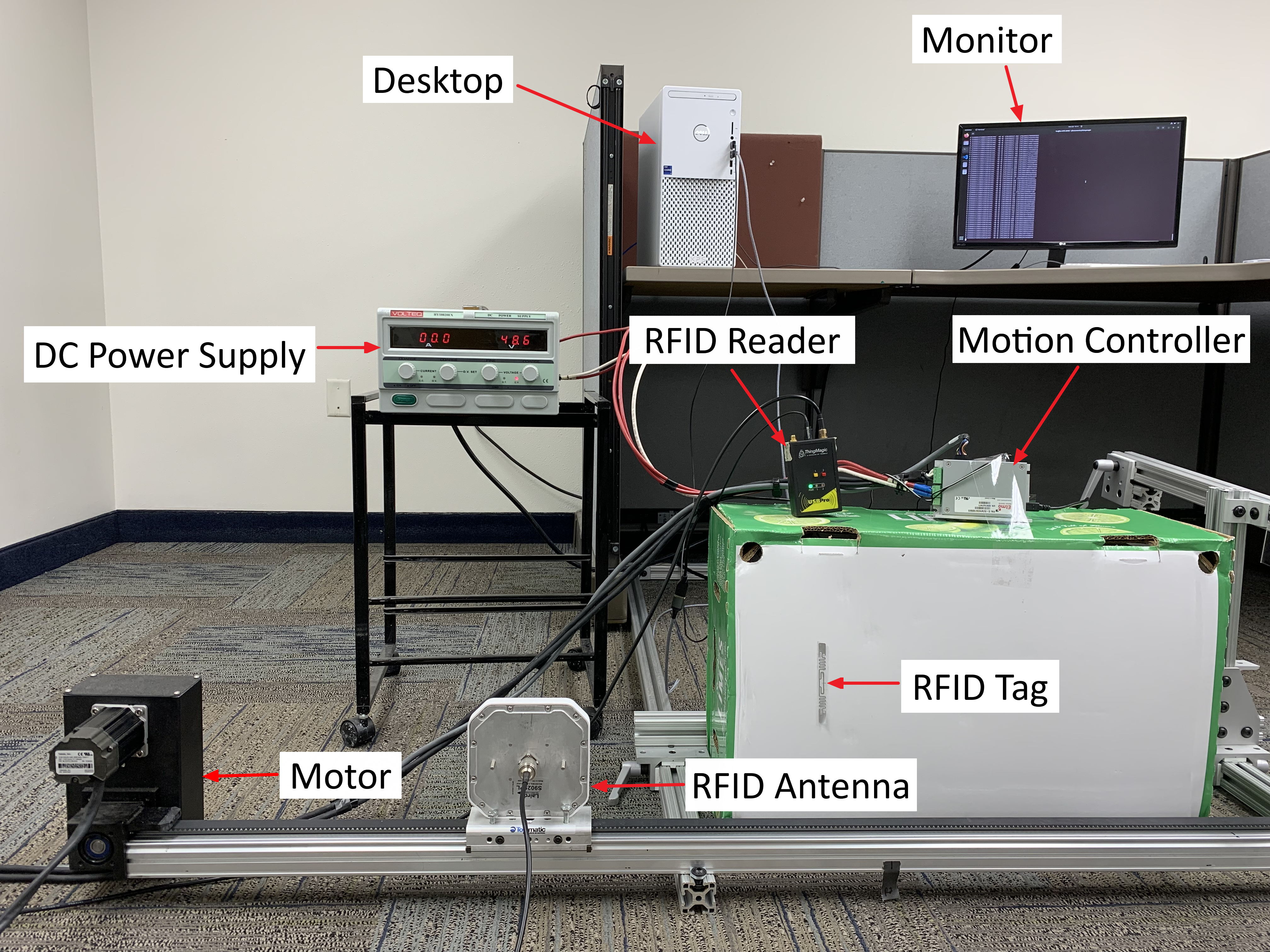}\label{fig:e1}}
  \hfill
  \subfloat[Field experiments]{\includegraphics[width=0.49\textwidth]{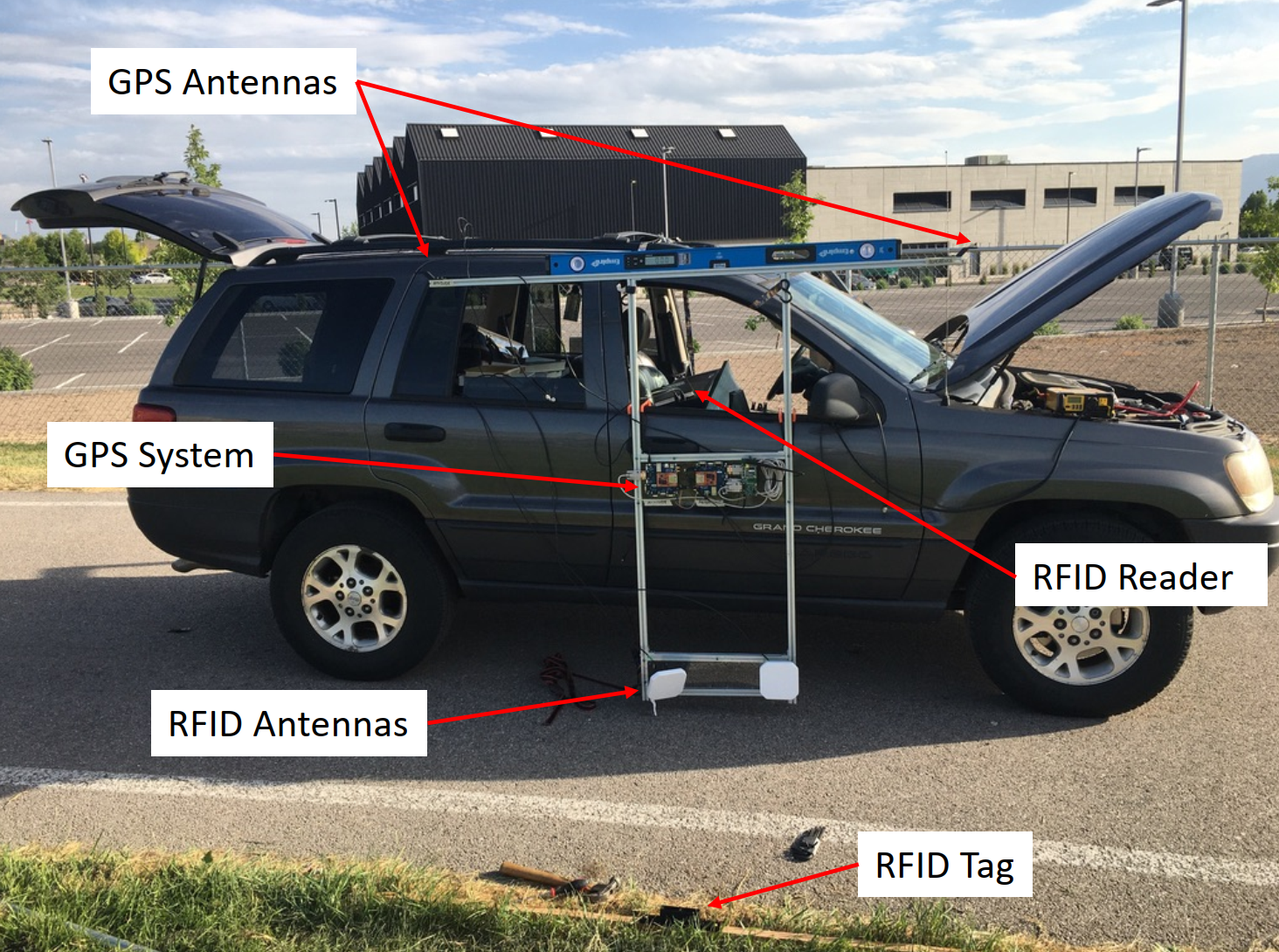}\label{fig:e2}}
  \caption{Lab and field experiments at Utah State University}
  \label{fig:experiments}
\end{figure}

Illustrated in Fig. \ref{fig:experiments}\subref{fig:e1}, we show the overall system configuration. Commercial RFID reader ThingMagic USBPro is used to obtain the transmit and receive phase difference ($\Psi_t - \Psi_R$).  The antenna is Laird S9025PL. The tag is ALN-9740 Squiggle from Alien Technology, fully compatible with the UHF EPC global Gen2 standard. Our setting assumes the tag location is known, and MLE will estimate the relative distance of the EV (reader).



\subsection{Indoor track evaluation results}

 We start with the indoor track scenario. Specifically, we build a rack system of 1.5m in length and place the RFID antenna on a movable cart. The antenna is fixed on the cart and can be precisely controlled by the Elmo motion controller and Teknic servo motor. We wrote the Python script to control the distance and the movement velocity. Therefore, the precise distance (ground truth) can be obtained as a reference. The RFID tag is placed on cardboard parallel to the track. The RFID antenna faces the tag direction while moving in the perpendicular direction. The perpendicular distance between the antenna and the tag is 0.2 m. The lab setup is shown in Fig. \ref{fig:experiments}\subref{fig:e1}.

To verify the effectiveness of the MLE algorithm, we control the cart starting from a known position and moving 0.005 m every step. At each step, the cart will stay for 0.1 s. Since the tag is fixed, we can calculate the closest Euclidean distance between the antenna and the tag at each step. Then, the theoretical phase can be calculated. In this process, the RFID reader will continuously read from the tag. The working frequency of the reader is set as 910 MHz with transmission power of 25 dBm. The ID, RSS, phase, and timestamp information will be collected for each read. The phase information of the tag will be sampled at each step. 

\begin{figure}[ht]
  \centering
  \subfloat[Likelihood result]{\includegraphics[width=0.48\textwidth]{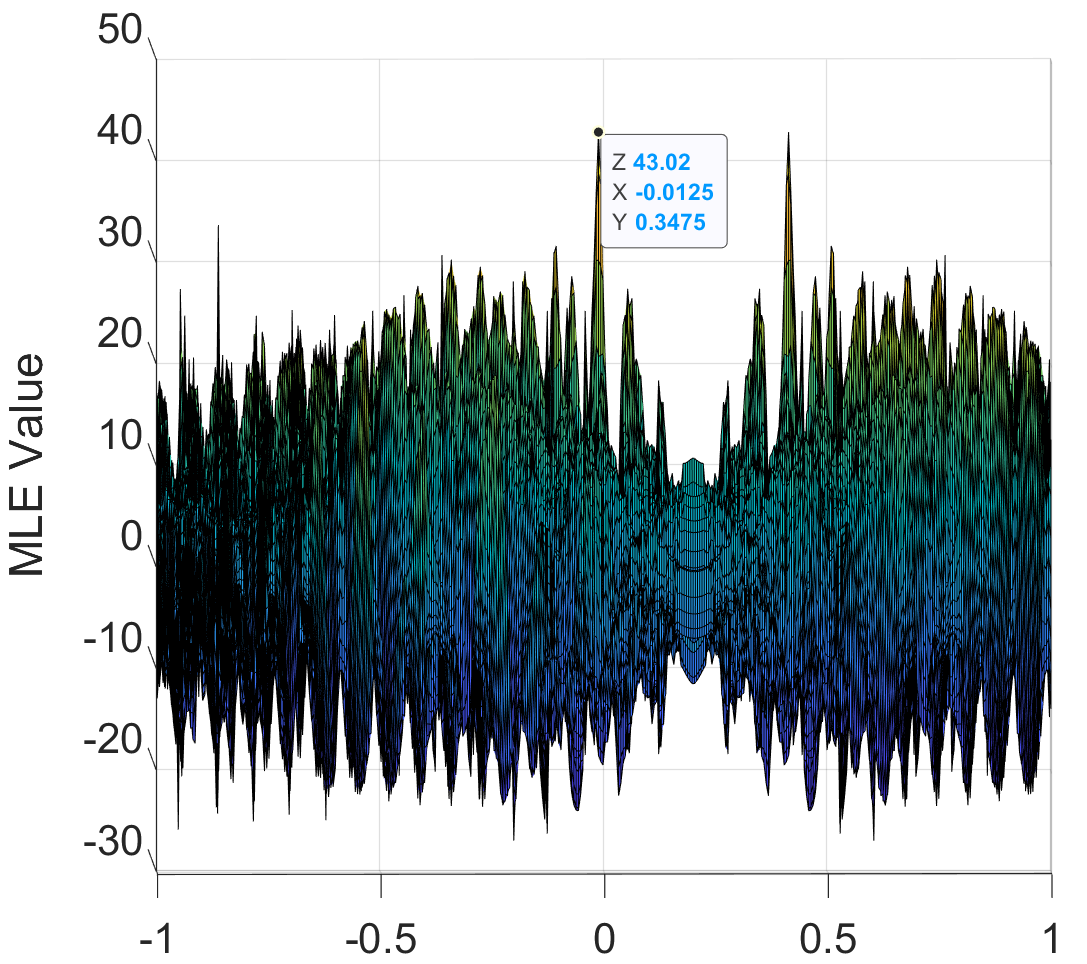}\label{fig:r1}}
  \hfill
  \subfloat[Phase result]{\includegraphics[width=0.49\textwidth]{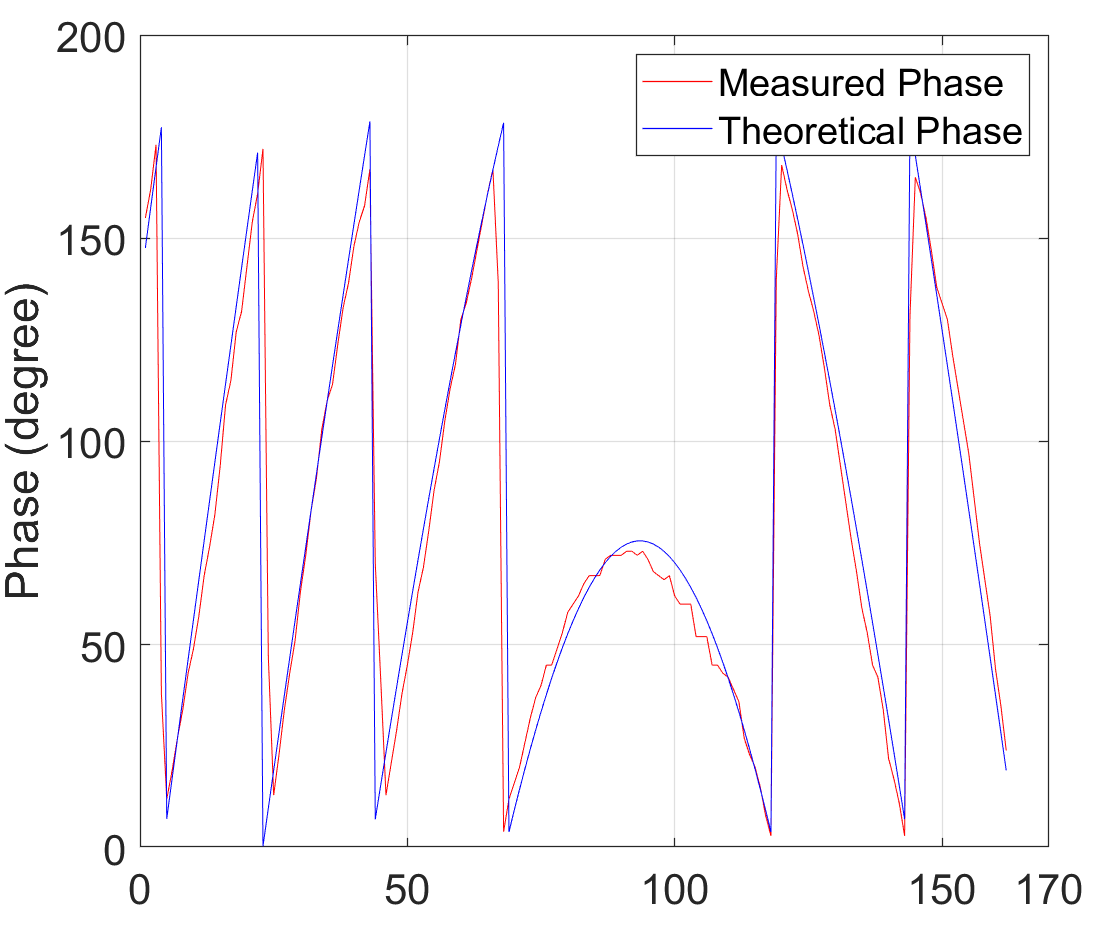}\label{fig:r2}}
  \caption{Lab experiments result}
  \label{fig:lab_result}
\end{figure}

The phase returned by the RFID reader is in the $(0, 180)$ degree range. The measured and grid phases are normalized in the $(-1,1)$ range to calculate the likelihood. The likelihood is calculated for each grid, and a likelihood map is constructed. The map has two peaks, as shown in Fig. \ref{fig:lab_result}\subref{fig:r1}. And the map is symmetric with the tag along the cart's moving direction. When the cart is known at the left side of the tag, then the other peak can be ignored. The measured phase and the theoretical phase starting from the maximum likelihood point are well aligned. The ground truth cart starting point is $(0, 0.38)$, and the MLE calculated result is $(-0.0125, 0.3475)$. The difference between the ground truth and the MLE result is 0.035 m or 3.5 cm. The error may come from the measurement or algorithm.

\subsection{Field experiments}
Fig. \ref{fig:experiments}\subref{fig:e2} shows the field experiments setup. It is difficult to know the Euclidean distance between the antenna and the tag since vehicle location is not fixed in a track. We use the dedicated GPS to provide position information to calculate the ``true'' phase. When the position of the RFID tag, the positions of GPS antennas, and the relative position of the RFID antennas are known, the distance between the RFID antenna and the RFID tag can be calculated. 

\begin{figure}
    \centering
    \includegraphics[width=0.8\linewidth]{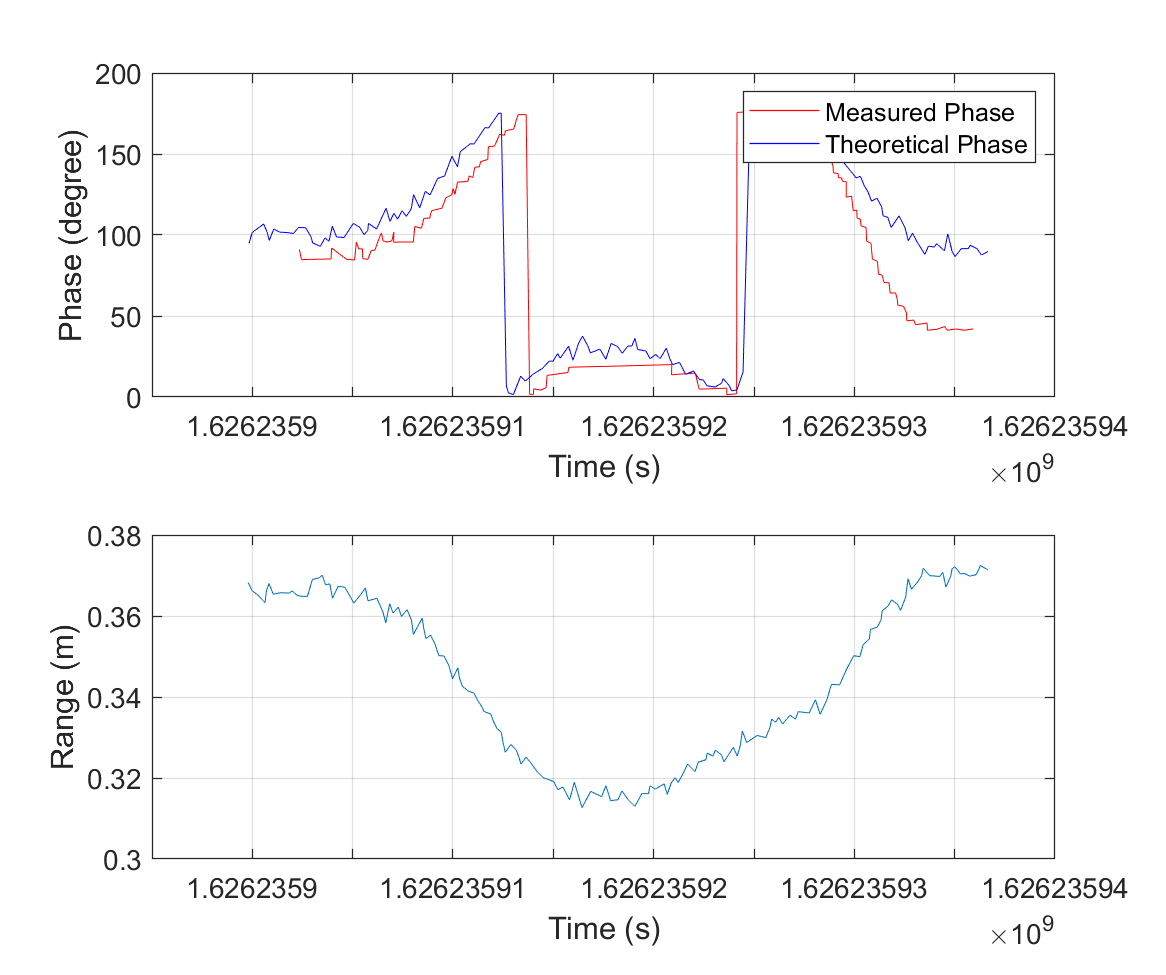}
    \caption{Field experiment result}
    \label{fig:field_test}
\end{figure}

As the vehicle passes by the RFID tag, the distance between the RFID antenna and the tag decreases and subsequently increases, as shown in Fig. \ref{fig:field_test}. With MLE, the calculated phases match well with the measured phase. The error, after GPS position correction, is around 5 cm. 

\section{Challenges and Future Opportunities}
In the previous sections, we have shown that an RFID-based coil alignment system is a valid approach. To unleash its full potential, additional challenges have to be addressed. 

\subsection{System implementation challenges}

The first challenge comes from the nature of wireless signals. In the system model and experiment, LoS is considered. But in real-world settings, NLoS is also very common. For example, if RFID tags or reader antennas are covered with any foreign object, their communication will be significantly impacted, leading to inaccurate phase measurement. Besides, the time-varying multipath effect will result in the signal arriving at the receiver via multiple reflections, creating additional phase ambiguity. This problem is more severe in EV scenarios since its metal surface makes strong signal reflection indistinguishable from the direct LOS path. Mitigating the effects of multipath requires sophisticated signal-processing techniques and algorithms. Designing and implementing these algorithms can be complex and require additional computational resources.

The second challenge revolves around  EV high mobility. 
The principle of a passive RFID tag suggests that it needs to be energized by the reader first and then send information back. That means the reader should keep in the operational zone long enough for these two steps to be completed.  While this usually only takes a few milliseconds, the short range of the RFID reading zone poses challenges in obtaining more reliable readings when passing one tag. Moreover, antenna and tag polarization changes frequently during EV movement, impacting the reading rate. As mentioned, the MLE algorithm needs many coherent phase values from one tag to resolve phase randomness and ambiguity. Additionally, the coherent phase will need EVs' precise motion vector (speed and time interval), which can be very challenging to obtain. In \cite{yang2014tagoram}, they proposed an augmented algorithm to perform tag localization and reader movement estimation, which can be a potential solution. However, this algorithm prefers slow motions.

The third challenge pertains to  signal interference. For smooth operation, many RFID tags should be deployed along the road of transmitter coils. At any given time, more than one tag may respond to the reader's query,  making it challenging for the reader to identify and communicate with individual tags. As mentioned before, localization is relative to each tag. Therefore, it is necessary to differentiate each tag and obtain their corresponding reading at a high sampling rate. 

The fourth challenge comes from integrating misalignment with real-time EV automatic control. The important observation is that misalignment detection and estimation from RFIDs are not the ultimate goal. Instead, when coil misalignment is detected and estimated, an active alignment system should work on dynamically adjusting the position and orientation of the EV. Ideally, the system should automatically control micro-movement ($\leq$ 10 cm) of EVs. This will require a feedback mechanism that can continuously calculate distance from RFID sensing and apply predictive control to anticipate future EV movement changes.  

The last challenge concerns standard and interoperability. As shown in section II, the current state-of-the-art coil alignment is sporadic, initiated by researchers with very different approaches. However, standardization and interoperability are critical for the large deployment of the DWPT coil alignment system. EV manufacturers and infrastructure providers should set standards in both hardware and software alignment systems so that their interoperability can be achieved. 

\subsection{ Future opportunities}
Although there exist challenges to the wide adoption of RFID coil alignment systems, as sensing and control technology evolve, we expect to see more enabling approaches that can tackle those challenges. 

The first opportunity lies in the ultra-wideband RFID. The simple design of the RFID tag enables versatile modifications. For example, energized signals from the reader can range from narrowband frequencies (tens to hundreds of KHz) to over 200 MHz. As shown in \cite{ma2017minding}, leveraging the tag's underlying physical properties, signal across bandwidth exceeding 220 MHz is emulated without modification to the tag. Wide bandwidth signals provide better resolution in time and frequency domains and can help mitigate multipath effects by offering better discrimination between direct and reflected signals. Therefore, \cite{ma2017minding} reports a sub-centimeter 3D localization accuracy can be achieved.  

The second opportunity entails improving RFID sensing accuracy with multi-dimensional sensing information. In most existing works, including ours, single-dimensional measurement is applied. In \cite{9796909}, they can obtain channel state information data with up to 150-dimensional samples. The key observation is that RFID tags can acknowledge single-tone energizing and multi-tone signals. In their experiments, 150 subcarriers are used, each spanning a 125 KHz narrowband. These fine-grained results and possible ultra-wideband extensions can lead to higher-resolution results, even without a precise MLE algorithm. 

The third enabler is efficient signal processing based on machine learning. In our MLE algorithm, one challenge is the real-time grid calculation. Advanced learning algorithms like convolutional neural networks (CNN) or recurrent neural networks (RNN) are particularly effective in handling time-series data. In the case of MLE, likelihood calculation, filtering, and feature extraction with those algorithms will lead to faster analysis, reduced latency, and more efficient handling of small grid sizes.  

Lastly, we envision a multi-sensor data fusion approach as a promising avenue. By combining sensory inputs from camera, RFID, and EV controllers, a more comprehensive understanding of EV-coil positions can be obtained. For example, the camera sensor can provide reliable data for the MLE grid search in ideal weather conditions. Meanwhile, EV controllers can feed real-time charging information that also reflects misalignment parameters. With the integration of learning algorithms, this approach is poised to provide increased accuracy and reliability.

\section{Conclusions}
DWPT has become a prominent technology for the wide adoption of EVs but faces many technical challenges. One of the main issues is the alignment between primary and secondary coils. This article first introduces existing misalignment techniques, including magnetic, vision, wireless sensing, machine learning, and voltage detection approaches, and then discusses their pros and cons. It is concluded that the required misalignment accuracy cannot be easily fulfilled with those techniques. We then introduce an RFID-based DWPT coil alignment system.  With coherent phase detection and MLE algorithm, the proposed system has the potential to achieve sub-10 cm accuracy. Our lab and experimental test verified the RFID system to estimate lateral and vertical misalignment. Lastly, we envisioned some challenges in deploying such systems and remedy solutions. The aim of this work is to provide an alternative and promising approach for enabling highly efficient DWPT as part of the EV future infrastructure support. 
\section{Acknowledge}
The authors would like to thank the ASPIRE center at Utah State University for the field test support. 

\bibliographystyle{IEEEtran}

\bibliography{lib}
\end{document}